\documentclass[10pt,lettersize,journal]{IEEEtran}
\ifCLASSINFOpdf

\else

\fi
\usepackage{amssymb,amsmath,epsfig}
\makeatletter
\def\maketag@@@#1{\hbox{\m@th\normalfont\normalsize#1}}
\makeatother
\usepackage{graphicx}
\usepackage{epstopdf}
\usepackage{microtype}
\usepackage{cite}
\usepackage{url}
\usepackage[none]{hyphenat}
\usepackage{booktabs}
\usepackage[font=footnotesize]{caption}
\usepackage[linesnumbered, ruled,vlined]{algorithm2e}
\usepackage{algpseudocode}
\usepackage{multirow}
\usepackage{nicefrac}
\everymath{\displaystyle}
\usepackage{color}
\usepackage[font=footnotesize]{caption}
\graphicspath{{Schematics/}} 
\usepackage{url}
\usepackage{hyperref}
\usepackage{dblfloatfix}  
\usepackage[labelformat=simple]{subcaption}

\usepackage{graphicx}
\usepackage{booktabs}
\usepackage[table,xcdraw]{xcolor}
\usepackage{lipsum}

\makeatletter
\def\blfootnote{\xdef\@thefnmark{}\@footnotetext}
\makeatother

\begin{document}

	\title{{Time Index Modulation-Driven Standalone RIS Mechanism for  Symbiotic Radio}}
	
	\author{M. Ertug Pihtili~\IEEEmembership{Student Member,~IEEE}, Mehmet C. Ilter~\IEEEmembership{Senior Member,~IEEE},  Ertugrul Basar~\IEEEmembership{Fellow,~IEEE}\vspace{-0.7cm}}
		
		\IEEEtitleabstractindextext{
		\begin{abstract}
		The rising demand for energy and spectrum resources in next-generation Internet-of-things (IoT) systems accounts for innovative modes of information and power transfer. One potential solution is to harness the active transmission capability of devices to facilitate data transmission and wireless energy harvesting (WEH) for backscatter communication so as to form a symbiotic radio (SR) environment in a mutualistic manner. Additionally, incorporating reconfigurable intelligent surfaces (RISs) into the SR environment can provide an additional link and enhance the reliability of backscatter communication, thereby reinforcing the symbiotic relationships between active and passive devices. This paper proposes a novel SR system where a standalone RIS sustains its functions through WEH based on a low-power RIS structure and establishes mutualistic symbiosis by utilizing a signal conveyed by the primary transmitter (PTx) to assist ongoing transmissions and convey information to the primary receiver (PRx). The PTx employs time index modulation (TIM) to transmit information to the PRx and power to the RIS and energy harvester (EH). A log-likelihood ratio (LLR)-based detector is presented to address challenges in the TIM scheme. Finally, the performance of the proposed scheme is investigated in terms of harvested direct current (DC) power at the RIS and EH, as well as the bit error rate (BER) at the PRx.
		\end{abstract}
				
		\begin{IEEEkeywords}
				Wireless energy harvesting, reconfigurable intelligent surfaces, symbiotic radio, index modulation.
		\end{IEEEkeywords}}
		\maketitle
		\IEEEdisplaynontitleabstractindextext
		
		\IEEEpeerreviewmaketitle
		\section{Introduction}

		\IEEEPARstart{T}{he} integration of wireless energy harvesting (WEH) into future Internet-of-things (IoT) architectures is increasingly vital as the coexistence between cellular and IoT devices demands additional energy resources to support the growing number of connected devices. In order to address rising demand, symbiotic relationships among cellular and IoT devices have been discovered, whereby the passive transmission capabilities of IoT devices leverage the power and spectrum resources of active transmission technology used by cellular devices. The concept, known as symbiotic radio (SR), has been introduced to enable energy- and spectrum-efficient communications within the network  \cite{Liang2020}.
		
		Similar to SR, IoT devices can utilize ambient backscatter communications (AmBC) by leveraging existing ambient RF signals in the environment. However, AmBC suffers from direct link interference from surrounding ambient RF signals, which limits detection performance at the target receiver \cite{long2020}. Contrary to AmBC, the SR concept forms a novel architecture in which active and passive systems rely on mutualistic communication through joint decoding at the receiver and an additional link provided by a backscatter device (BD) rather than parasitic communication, which is a drawback of AmBC \cite{Liang2020}. In the SR framework, the primary transmitter (PTx) shares its energy and spectrum resources with the secondary transmitter (STx). Meanwhile, STx operates as a passive device that also transmits its information to primary receiver (PRx) by manipulating the resources of PTx. To establish mutualistic symbiosis, the PTx maintains a higher transmission rate than the PRx, which means the symbol duration of the STx is longer than that of PTx \cite{long2020}. In \cite{Zhang2022}, the authors reveal that the mutualistic condition depends on the average strength of direct and backscatter links, and \cite{Liang2020} mentions that the backscatter link can be improved by utilizing reconfigurable intelligent surfaces (RISs). The RIS-aided SR system has recently introduced in \cite{Yang2024}, where the RIS performs two functions simultaneously: assisting PTx transmission and transmitting its information by orchestrating the phase shifts on the incident signals.
		
		Specifically, RISs are passive devices consisting of $N$ unit cells (UCs) that operate as reflectors to manipulate the wireless channel by tuning the phase of the reflected signal \cite{BasarRIS}. RIS UCs are tunable devices equipped with PIN diodes, varactors, and RF switches, that control the wireless medium and enable the reflection function. Nevertheless, UCs can also be operated as absorbers for WEH on the RIS. In \cite{Petrou2022}, a novel RIS controller architecture, called integrated architecture, is discussed to form a low-power RIS setup. Based on this integrated architecture, WEH protocols for self-sustainable operations are explored in \cite{Ntontin2024}, assuming perfect absorbers with no reflected signals caused by the RIS. To achieve a self-sustainable architecture, the harvested direct current (DC) power on the RIS should be at least equal to its total power consumption. Although the integrated architecture provides lower power consumption than conventional RIS controllers, the type of UCs can make self-sustainable operations infeasible if varactor-based UCs are employed, as they are driven by power-hungry digital-to-analog converters (DACs). The authors examined the power consumption of various UCs and claimed that it is possible to design an RF switch-based RIS without DACs \cite{wang2024reconfigurable}. An RF switch-based RIS prototype was designed, demonstrating that by connecting one port of the RF switch to the absorber, WEH can be achieved on RIS \cite{rossanese2022}.
		
		{\blfootnote{{\textit{Notations:} Italic letters $x$ denote scalar values, while lowercase $\mathbf{x}$ and uppercase $\mathbf{X}$ in boldface represent vectors and matrices, respectively.  The superscripts  $(\cdot)^T$ and $(\cdot)^H$ denote the transpose and the Hermitian transpose operations, respectively. $\angle\cdot$, $\lvert\cdot\rvert$, and $\lVert\cdot\rVert$ denote the angle, absolute value, and the norm of a vector/matrix, respectively. $\log_a(\cdot)$ represents the logarithm with base $a$. The operator $\mathbb{E}[\cdot]$ indicates averaging/statistical expectation. $\Pr(\cdot)$ represents the probability of an event. $\lceil \cdot \rceil$ and $\lfloor \cdot \rfloor$ denote the ceil and floor operations, respectively. The symbol $j$ represents the imaginary unit $\sqrt{-1}$, while $I$ and $Q$ denote a complex symbol's in-phase and quadrature parts, respectively. $\mathcal{CN}(\mu, \sigma^2)$ represents a circularly symmetric complex Gaussian (CSCG) random variable with mean $\mu$ and variance $\sigma^2$. The set of complex matrices of dimensions $m \times n$ is denoted by $\mathbb{C}^{m \times n}$, the set of non-negative integers of dimensions $m \times n$ is denoted by $\mathbb{N}^{m \times n}$, and $\mathbb{Z}^{+}$ denotes the set of integers greater than zero. The binomial coefficient is denoted by $C(n, k)$. The $\mathbf{0}_{n \times m}$ denotes the all-zero matrix of size $n \times m$.}}}
		
		To address the increasing power demand, simultaneous wireless information and power transfer (SWIPT) emerges as a candidate for next-generation green communication systems, wherein power and information transmission are handled separately across various domains: power (power splitting), time (time switching), and space (antenna splitting) \cite{Ponnimbaduge2018}. An alternative approach to SWIPT, called Information Harvesting (IH), has been proposed to synergistically transfer information and power to IoT devices by exploiting a far-field RF power transfer mechanism for data transmission through index modulation (IM) while avoiding signal splitting into different domains \cite{Ilter2022}.
	
		IM can map information bits by altering the on/off status of transmission entities in a communication system, such as transmit antennas and subcarriers \cite{BasarIM2017}. This distinctive feature enables the integrated transfer of power and information by means of IM. To facilitate data transfer and wireless energy harvesting (WEH) efficiently, a time index modulation (TIM) technique that employs time slots to convey additional information is presented in \cite{Zhao2023}. In \cite{Basar2013}, an orthogonal frequency division multiplexing (OFDM)-IM scheme is proposed by inactivating some subcarriers, and its performance under the low-complexity log-likelihood ratio (LLR) detector is investigated. Furthermore, \cite{mao2016} presents a dual-mode OFDM-IM system by replacing inactive subcarriers with a different constellation set from those used in active ones. In \cite{Tugberk2024}, the error performance of several OFDM-IM-based schemes is compared using an LLR-based detector. Building upon the concept introduced in \cite{Ilter2022}, a novel OFDM-based information harvesting concept is proposed in \cite{Ilter2024}, where OFDM-IM is applied to an ongoing wireless power transfer mechanism. \cite{Lin2021} proposes an RIS-based IM scheme in which the indices of the activated group of RIS UCs are utilized to convey environmental data collected by RIS. To prevent power loss caused by inactivating a group of RIS UCs, partitioning the RIS UCs into two orthogonal groups is proposed in \cite{Lin2022}.
		
		In this paper, we present a novel TIM-driven SR environment empowered by a standalone RIS architecture. This architecture utilizes an integrated controller to reduce the power consumption of the RIS and employs 2-bit resolution RF switches as UCs to eliminate power-hungry DACs, thereby promoting sustained functionality through WEH. Consequently, the RIS operates as a BD in the SR environment by initially applying WEH, subsequently assisting PTx transmission strategy and conveying RIS information by inducing phase shifts selected from a discrete set. Additionally, PTx implements the TIM scheme to power the RIS and EH while ensuring reliable communication with the PRx. Moreover, we propose an LLR-based detector tailored for the TIM-based SR scheme within our framework to address decoding challenges.

		\section{Symbiotic Radio Model}
\begin{figure}[t!]
			\centering
			\includegraphics[width=0.5\textwidth]{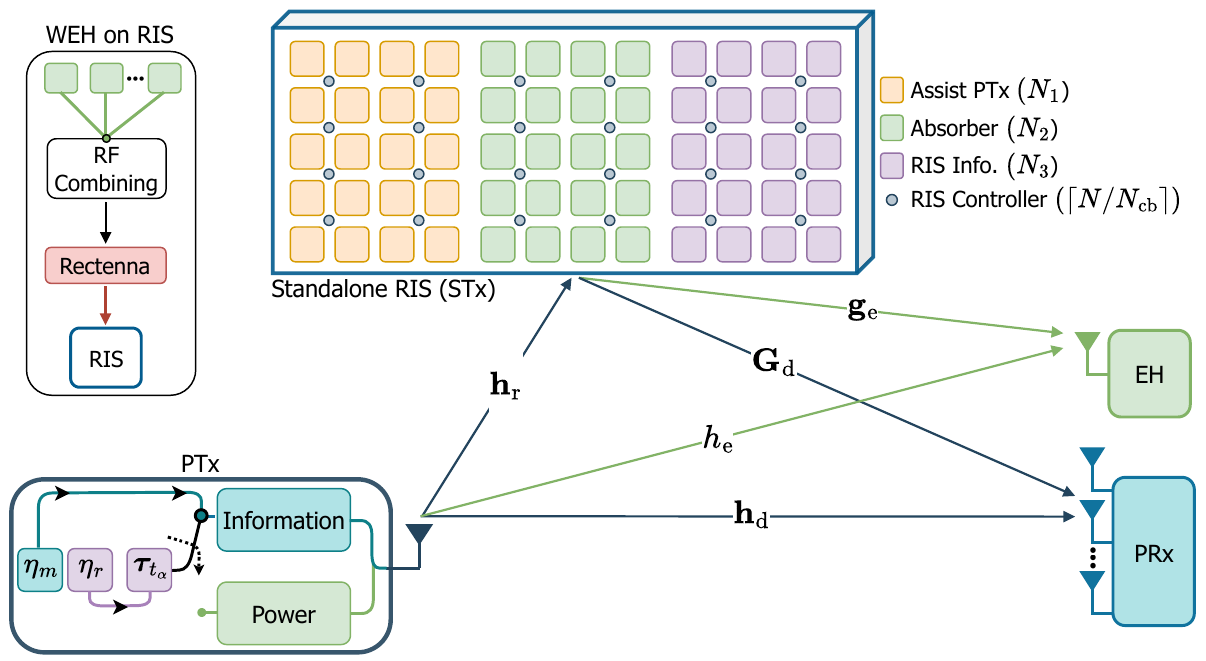}
			\caption{System model of TIM-aided RIS-SR scheme.}
			\label{fig:TIM_SR_RIS_Sys}
		\end{figure}

		In the proposed scheme, the primary system, which consists of a single antenna PTx and an $M_\text{R}$ antenna PRx, relies on active RF transmission technology. The secondary system comprises a standalone RIS with $N = N_1 + N_2 + N_3$ UCs and a single antenna EH, both of which are passive IoT devices. These passive IoT devices harness the power and spectrum resources of the primary system to sustain their operations such that the primary and secondary systems rely on symbiotic relationships \cite{Liang2020}. In Fig. \ref{fig:TIM_SR_RIS_Sys}, PTx leverages the TIM technique to send information to the PRx and power to the EH through the direct link and the link empowered by RIS. Therein, the RIS is partitioned into three groups: the first group, with $N_1$ UCs, assists the PTx transmission; the second group uses $N_2$ UCs as absorbers to ensure the RIS's self-sustainability through WEH by exploiting the PTx signal; and $N_3$ UCs in the third group are dedicated to conveying the RIS information by inducing phase shifts in the reflected signal.
		The PTx-RIS and RIS-PRx channels are defined as $\mathbf{h}_r  \triangleq\left[{\mathbf{h}}_{r,1}, {\mathbf{h}}_{r,2}, {\mathbf{h}}_{r,3}\right]^T \in  \mathbb{C}^{N \times 1}$ and $\mathbf{G}_d \triangleq\left[{\mathbf{G}}_{d,1}, {\mathbf{G}}_{d,2}, {\mathbf{G}}_{d,3}\right]  \in  \mathbb{C}^{N_\text{R} \times N}$, respectively, where subscripts $l = \{1, 2, 3\}$ associated with the RIS groups with $\{N_1, N_2, N_3\}$ UCs. All links follow the Rician channel model, and  $v$-th entry of PTx-PRx, and  $(v,n)$-th entry of RIS-PRx link can be written as:
		\begin{align}
			h_d^v &=  \sqrt{\nu_p}\left(\sqrt{\frac{\kappa_p}{\kappa_p+1}} {h}_{d,\mathrm{LoS}}^n+\sqrt{\frac{1}{\kappa_p+1}} h_{d,\mathrm{NLoS}}^n\right), \\ 
			g_d^{v,n} &= \sqrt{\nu_p}\left(\sqrt{\frac{\kappa_p}{\kappa_p+1}} g_{d,\mathrm{LoS}}^{n,v}+\sqrt{\frac{1}{\kappa_p+1}} g_{d,\mathrm{NLoS}}^{n,v}\right),
			\label{eq:f_CH}
		\end{align}
		where $v \in \{1, 2, \hdots, M_R\}$ and $n \in \{1, 2, \hdots, N\}$. Herein, $\nu_p$ and $\kappa_p \geq 0$ for ${p} \in \{h_d, g_d\}$ represent the path loss and the Rician factors, respectively. The terms ${h}_{d,\mathrm{LoS}}$, and $g_{d,\mathrm{LoS}}^{v,n}$ are the deterministic line-of-sight (LoS) components  with $\lvert{h}^v_{d,\mathrm{LoS}}\rvert = \lvert g_{d,\mathrm{LoS}}^{v,n}\rvert = 1$ and $ {h}^v_{d,\mathrm{NLoS}}$, and  $ g_{d,\mathrm{NLoS}}^{v,n}$ represent non-LoS components in which each entry is independent and identically distributed (i.i.d.) according to $\mathcal{CN}\left ( 0,1 \right )$. The definitions of the remaining channels follow similar principles and can be derived analogously.
		
		\subsection{Primary transmission model}
		
		The PTx utilizes the TIM scheme to convey information to the PRx and power to the EH over direct and RIS links. Hence, the PTx also sends bits by indexing information and power transmission stages into distinct time slots, allowing deterministic power signals to participate in information transmission. Thus, there is an ongoing information transmission even when the PTx switches to a power stage.
	
		\subsubsection{Information Stage}
		In the information stage, $L$ information signals $x_{i} \in \mathcal{M}$ with a power of $P_L$ are drawn from $M$-PSK/QAM constellation points, where $\mathcal{M} = \{ x_{i} : 1 \leq i \leq M \}$. Therefore, $\eta_m = L \log_2(M)$ bits can be sent over the total block length. By leveraging TIM to alter the transmission ordering of the $L$ information signals, $r = C(K, L)$ distinct mappings of $L$ information signals can be achieved. Thus, additional $\eta_r = \left\lfloor \log_2(r) \right\rfloor$ bits are conveyed, such that $\eta = \eta_m + \eta_r$ bits are transmitted over one block duration that spans $K$ transmission stages.
		
		According to $\eta_r$ bits, the time indices of $L$ information signals are chosen from the set for each block duration, which is given by
		 \begin{equation}
		 	\mathcal{T} = \begin{Bmatrix} \mathbf{t}_1& \mathbf{t}_2& \hdots& \mathbf{t}_{2^{\eta_r}} \end{Bmatrix},
		 \end{equation}  
		in which there are $2^{\eta_r}$ possible legitimate time-index selections out of $r$. After deciding on $\mathbf{t}_\alpha \in \mathbb{Z}^+$, where $\alpha \in \{1, 2, \hdots, 2^{\eta_r}\}$, the time-index vector is defined as $\boldsymbol{\tau}_{\mathbf{t}_\alpha} = \begin{bmatrix} \tau_1 & \tau_2 & \hdots & \tau_K \end{bmatrix}^T \in \mathbb{N}^{K \times 1}$, and $\tau_k$ is set to 1 at every information transmission stage, where $k = 1, 2, \hdots, K$.
		\subsubsection{Power Stage}
		The remaining $(K-L)$ time slots are used for deterministic power signal $\omega = \omega_I + j\omega_Q$ with a power of $\lvert \omega\rvert^2 = P_H$, where $P_H \geq P_L$, and $\tau_k$ is set to 0 at each power transmission stage.
		
		By following the construction of $\boldsymbol{\tau}_{\mathbf{t}_\alpha}$, the $\eta_m$ bits are mapped to the $M$-PSK/QAM symbols, where each symbol carries $\eta_m / L$ bits. The mapping is done considering the legitimate signal set $\mathcal{S}$, which has a length of $M^L$ and is defined as follows:
		\begin{equation}
			\mathcal{S} =
			\begin{Bmatrix}
				\mathbf{x}_1&  \mathbf{x}_2 & \mathbf{x}_3& \cdots & \mathbf{x}_{M^L}
			\end{Bmatrix}.
		\end{equation}
			\begin{figure}[t!]
			\centering
			\includegraphics[width=1\columnwidth]{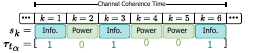}
			\caption{PTx transmission strategy.}
			\label{fig:TIM_RIS_SR_TxArch}
		\end{figure}
		Therein, $\mathbf{x}_n\in\mathcal{S}$, where $n \in \{1, 2, \hdots, M^L\}$, represents a vector of length $L$, $\mathbf{x}_n = \begin{bmatrix} {x}_{i,1}^n& {x}_{i,2}^n& \cdots&  {x}_{i,L}^n\end{bmatrix}^T$ in which, $P_L = \mathbb{E}[|x_{i}|^2]$ is normalized to unit power.
		Consequently, the elements of $\mathbf{x}_n$, namely the symbols, are mapped into the time slots based on $\boldsymbol{\tau}_{\mathbf{t}_\alpha}$ to initialize the transmission stage. Accordingly, the signal $s_k \in \{\mathcal{S}, \omega\}$ is transmitted  at each time instant $k$.

		In Fig. \ref{fig:TIM_RIS_SR_TxArch}, the transmission strategy of PTx is illustrated for $K = 6$ and $L = 3$. The total block duration is assumed to be less than the channel coherence time, ensuring that the channel coefficients remain unchanged for a block of length $K$, though they might change from one block to another. Herein, PTx alternates between information and power stages according to $\boldsymbol{\tau}_{\mathbf{t}_\alpha} = \begin{bmatrix} 1 & 0 & 1& 0 &0 &1 \end{bmatrix}$, where $\mathbf{t}_\alpha = \begin{bmatrix} 1 & 3 & 6 \end{bmatrix}$, indicating that the locations of the information and power stages in time domain carry additional information bits.
		
		\subsection{RIS Absorption Function and WEH}
		
		The absorbing function of UCs enables RIS operations through WEH without requiring an external battery or power grid. Hence, RF energy harvesting on the RIS can be achieved through RF switches since they can act as both reflectors and absorbers \cite{wang2024reconfigurable}. Accordingly, RF switches are utilized in RIS as UCs, with $N_2$ UCs serving as perfect absorbers and the remaining $N - N_2$ UCs functioning as perfect reflectors. The RF switch allocates one output port for an impedance-matching component, which absorbs the energy of incoming signals instead of reflecting them. The absorbed signals are fed to the RF combining network, which combines the signals in the RF domain \cite{BrunoRF} and outputs them to the nonlinear rectenna as input for obtaining harvested DC power such that the rectenna's input at time instant $k$ is mathematically given by
		\begin{equation}
			\mathrm{Q}_k = \left|\sum_{n_2 = N_1 + 1}^{N_2}{h}_{r,2}^{n_2}s_k \right|^2.
			\label{eq:Avg_Harvested}
		\end{equation}
			
		The constant-linear-constant (CLC) model is employed to capture the practical characteristics of a rectenna,  including turn-on sensitivity and saturation. \cite{Moon2021}. In the CLC rectenna model, the harvested DC power, $\mathcal{P}_{\text{DC}}(\mathrm{Q})$ is expressed as follows: 
		\begin{equation}
			\mathcal{P}_{\text{DC}}(\mathrm{Q}_k) = \begin{cases}0, & \mathrm{Q}_k \in [0, \mathcal{P}_{\text{on}})  \\ \rho(\mathrm{Q}_k - \mathcal{P}_{\text{on}}), & \mathrm{Q}_k \in [\mathcal{P}_{\text{on}}, \mathcal{P}_{\text{sat}})
				\\ \rho(\mathcal{P}_{\text{sat}} - \mathcal{P}_{\text{on}}), & \mathrm{Q}_k \in [\mathcal{P}_{\text{sat}}, \infty) \end{cases}.
			\label{eq:CLC}
		\end{equation}
		Here, $\rho$ denotes the WEH efficiency, and $\mathrm{Q}_k$ represents the input power to the rectifier at time instant $k$. The turn-on sensitivity $\mathcal{P}_{\text{on}}$ is the minimum received DC power required to activate the rectenna for RF-to-DC conversion. Saturation occurs when the rectenna reaches its maximum power harvesting capacity.
		The maximum harvested DC power is expressed as $\mathcal{P}_{\text{max}} = \eta(\mathcal{P}_{\text{sat}} - \mathcal{P}_{\text{on}})$. Since the input power of the rectifier circuit varies depending on the 	specific application, $\mathcal{P}_{\text{sat}}$ and $\mathcal{P}_{\text{on}}$ should be adjusted according to the input power level. When the input power to the rectifier circuit is high, increasing $\mathcal{P}_{\text{sat}}$ can optimize the benefits of the high input power.

		Due to the fact that the RIS relies on WEH through its $N_2$ absorbers to serve its functions, the harvested DC power $\mathcal{P}_{\text{DC}}(\mathrm{Q}^\text{RIS}_k)$ should be sufficient to maintain the total power consumption of the RIS. The total power consumption required to operate the RIS consists of two parts: the static power consumption caused by the RIS controller and the dynamic power consumption generated by the RIS UCs. Therefore, the total power consumption modeling of RIS can be expressed as follows:
		\begin{equation}
			P_{\text{RIS}} = P_{\text{stc}} \, + \, P_{\text{dyn}}
			\label{eq:RISPC}
		\end{equation}
		where $P_{\text{stc}}$, and $P_{\text{dyn}}$ refer to the static and dynamic power consumption of RIS, respectively. $P_{\text{stc}}$ consists of the power consumption of the control board, $P_{\text{cb}}$, which has a constant value, and the power consumption of the drive circuits, $P_{\text{drv}}$, which varies with the type of UC driving mechanism. Therefore, various UCs, such as varactors and RF switches, exhibit distinct power consumption characteristics \cite{Petrou2022}. Since varactor-based UCs require power-hungry DACs to achieve continuous phase shifts, they lead to increased static power consumption. As $N$ increases, $P_{\text{stc}}$ tends to dominate $P_{\text{dyn}}$, making the self-sustainable condition challenging \cite{Ntontin2024}. To accomplish a standalone RIS architecture, one can utilize RF switches directly driven by RIS controllers without DACs \cite{wang2024reconfigurable}. Additionally, an integrated architecture for the RIS controller is a favorable approach to constructing a standalone RIS, as it consumes very low power during static conditions compared to conventional RIS controllers \cite{Petrou2022}. Thus, RF switches combined with an integrated architecture could offer a potential solution to meet the standalone RIS condition despite the discrete phase shift capability of RIS compared to the continuous phase shifts provided by varactor-based UCs. The power consumption of the integrated architecture for varactors and RF switches is given by, respectively,
		\begin{equation}
			 {P}_{\text{RIS}} = \begin{cases} \left\lceil N/N_{\text{cb}}\right\rceil P_{\text{cb}} + N \left(2P_{\text{drv}} + P_{\text{varactor}}\right), & \text{for varactor}	\\	
			\left\lceil N/N_{\text{cb}}\right\rceil P_{\text{cb}} + N P_{\text{switch}}, & \text{for RF switch} \end{cases},\label{eq:PtASICRF} 
		\end{equation}
		where $N_{\text{cb}}$ is the number of UCs controlled by one RIS controller, and $P_{\text{dyn}} \in \{P_{\text{switch}}, P_{\text{varactor}}\}$. The condition that verifies the standalone RIS operation can be given as follows:
		\begin{equation}
			\frac{1}{K}\sum_{k=1}^{K}\mathcal{P}_{\text{DC}}(\mathrm{Q}^\text{RIS}_k) \geq P_{\text{RIS}},
			\label{eq:zeRIS}
		\end{equation}
		which implies that the average harvested DC power over a block of length $K$ should be at least equal to or greater than the power consumption of the RIS.
		
		\subsection{RIS-enabled passive transmission model}
		During the information stage, the RIS performs two functions: it assists PTx transmission using $N_1$ UCs and transmits RIS information through $N_3$ UCs by applying a phase shift to the signal. Since RF switches are utilized, they can induce discrete phase shifts based on the $b$-bit resolution from a discrete set, i.e., $\phi_{l} \in$ {{$\mathcal{J}=\left\{0, {2 \pi}\slash {B}, \ldots, {(B-1)\pi}\slash{B}\right\}$}}, where $\phi_l$ represents the common phase shift within the $l$-th group. The RIS can generate $B = 2^b - 1$ different phase shift levels, as one output port of the RF switches is dedicated to an absorber. Once designed, these phase shifts remain invariant across all channel realizations in the RF switch-based RIS architecture. If a 2-bit resolution RF switch is employed, the possible phase shifts can be expressed as follows:
		\begin{equation}
		\phi_l \in \begin{Bmatrix}
			\phi_1 &  \phi_2 & \phi_p
		\end{Bmatrix}.
		\label{eq:RIS_Phases}
		\end{equation} 
		
		Considering the discrete set in \eqref{eq:RIS_Phases}, in the power stage, the RIS UCs configure the phases of $N - N_2 = N_1 + N_2$ UCs to $\phi_p$ for power transmission. In the information stage, the RIS selects phase shifts from the set $\mathcal{J}^\prime$, where $\mathcal{J}^\prime = \begin{Bmatrix} \theta_c : 1 \leq c \leq J \end{Bmatrix}$, to represent one bit of information based on the two possible phase options $\theta_c \in \{\phi_1, \phi_2\}$; therefore, the modulation order of RIS information is $J = 2$. The first group of $N_1$ UCs can alter their phase shifts at each information transmission slot by selecting the closest phase from the set $\mathcal{J}^\prime$ to assist PTx transmission. Meanwhile, to achieve mutualistic SR \cite{Liang2020}, the RIS aligns the third group of $N_3$ UCs to an invariant phase shift across $L$ information slots during a block of length $K$. Therefore, the symbol period of RIS information is larger than that of PTx, provided that $L > 1$, and the parameter $\delta = L$ represents the spreading gain that affects the mutualistic symbiosis, and increasing $\delta$ improves the detection performance of RIS information at PRx \cite{Zhang2022}.

		Let $\psi_l= \beta_le^{-j\phi_l}$ denote the reflect beamforming coefficient at the $l$-th RIS group in which $\beta_l \in[0,1]$ is the common reflection amplitude within the $l$-th group and $\boldsymbol{\Psi}_k=\begin{bmatrix}\psi_1&\psi_2 &\psi_3	\end{bmatrix}^T$ denote the reflection vector. The received signal at PRx at time instant $k$ is written as:	
		\begin{align}
			\mathbf{y}_k =  &\mathbf{h}_d s_k +{\sum_{n_1 = 1}^{N_1}\mathbf{g}_{d,1}^{n_1} e^{-j\phi_l}{h}_{r,1}^{n_1}s_k} \nonumber\\
			&+ \sum_{n_3 = N_2+1}^{N_3}\mathbf{g}_{d,3}^{n_3}  e^{-j\phi_l}{h}_{r,3}^{n_3} s_k+\mathbf{z}_k.
			\label{eq:y}
		\end{align}
		
		Herein, the reflection amplitudes are fixed to $\beta_l = 1$ for the first and third groups to maximize the reflected power, while $\beta_2 = 0$ since the second group absorbs the signal such that  $\psi_2=0$.
		Additionally, $\mathbf{z}_k \in \mathbb{C}^{M_{\text{R}} \times 1}$ denotes the additive white Gaussian noise (AWGN) sample vector, where each element ${z}_k \sim \mathcal{CN}\left(0, \sigma^2\right)$ has a zero mean and variance $\sigma^2$, added to the signal at each information stage.  Defining $\underline{\mathbf{f}}  \triangleq \begin{bmatrix}\mathbf{G}_{r,1}\mathbf{h}_{r,1}&\mathbf{G}_{r,2}\mathbf{h}_{r,2}&\mathbf{G}_{r,3}\mathbf{h}_{r,3} 	\end{bmatrix}  \in  \mathbb{C}^{N_\text{R} \times 3}$ as PTx-RIS-PRx cascaded channel matrix, the \eqref{eq:y} can be rewritten as:
		\begin{equation}
			\mathbf{y}_k =  \mathbf{h}_d s_k + \underline{\mathbf{f}}\boldsymbol{\Psi}_ks_k +\mathbf{z}_k.
		\end{equation}
		
	 	Similarly, the received signal at EH is given by
		\begin{equation}
			{\varepsilon}_k =  h_e s_k + \underline{\mathbf{v}}\boldsymbol{\Psi}_ks_k,
		\end{equation}
		where $\underline{\mathbf{v}}  \triangleq \begin{bmatrix}\mathbf{h}_{r,1}\mathbf{g}_{e,1}  &\mathbf{h}_{r,2}\mathbf{g}_{e,2}&\mathbf{h}_{r,3}\mathbf{g}_{e,3}	\end{bmatrix}  \in  \mathbb{C}^{1 \times 3}$ denotes the PTx-RIS-EH cascaded channel matrix and AWGN term is omitted since it is too low to be harvested. Therein, $\mathrm{Q}_\text{EH} = |\varepsilon_k|^2$ is input for the CLC rectenna defined in \eqref{eq:CLC} and the average harvested DC power can be calculated by putting  $\mathrm{Q}^\text{EH}_k$ into \eqref{eq:zeRIS}.
		
		The cascaded channel of the PTx-RIS-PRx (EH) links can be estimated using the method in \cite{Jensen2020}. Subsequently, we assume that PTx perfectly knows the cascaded channel and direct link channel. Therefore, PTx can transmit the phase knowledge and the corresponding transmission stage to the RIS prior to each transmission by determining the phase shift of the first group through $\psi_1 = \arg \min_{\theta_c \in \mathcal{J}^\prime} \lvert \theta_c - \mathbb{E}[(\angle \mathbf{G}_{r,1} \mathbf{h}_{r,1} + \angle h_d)] \rvert^2$. Accordingly, the RIS sets the phase of $N_1$ UCs and arranges the phase of $N_3$ UCs at time instant $k$.
		
		\section{SR-based Decoder}
		
		The PRx task involves jointly estimating the time-index vector $\boldsymbol{\tau}_{\mathbf{t}_\alpha}$ as well as the information of PTx and RIS by considering all possible time-index selections and corresponding symbols from the sets $\mathcal{S}$ and $\mathcal{J}^{\prime}$, respectively. To do so, PRx employs a maximum likelihood (ML) detector and minimizes the following metric:
		\begin{equation}
			\left(\boldsymbol{\hat{\tau}_{\mathbf{t}_\alpha}}, \mathbf{\hat{x}}_n, \hat{\psi}_3\right) = \arg \min_{\substack{{\mathbf{x}_n \in \mathcal{S}}\\ {{\mathbf{t}_\alpha}\in \mathcal{T}}\\ {\psi_3 \in \mathcal{J^{\prime}}}}}  \sum_{k=1}^{L}\lVert\mathbf{y}_k-\mathbf{h}_d s_k + \underline{\mathbf{f}}\boldsymbol{\Psi}_ks_k\rVert^2.
		\end{equation}
			\begin{algorithm}[t!] 
			\label{TIM_LLR}
			\SetAlgoLined
			\caption{LLR-Based Detector for TIM} 
			\DontPrintSemicolon
			\SetKwInOut{Parameter}{Require}
			\SetKwInOut{Input}{Ensure}
			\Parameter{$\mathbf{y}_{k}$, $\mathbf{h}_d$, $\underline{\mathbf{f}}$, $\boldsymbol{\Lambda}_c$, $\boldsymbol{\Omega}$, $\mathcal{J}^\prime$, $\mathcal{M}$, $\sigma^2$, $K$, $L$ }
			\Input{$\vartheta_k$ is the LLR of $k$-th time instant}
			\For {$k = 1$ \textbf{to} $K$}
			{	 
				$\Delta_i$ = $-\frac{1}{\sigma^2} \lVert\mathbf{y}_k-\mathbf{h}_dx_1 + \underline{\mathbf{f}}\boldsymbol{\Lambda}_1x_1\rVert^2 $\;	 
				$\Delta_p$ = $-\lVert\mathbf{y}_k-\mathbf{h}_d \omega + \underline{\mathbf{f}}\boldsymbol{\Omega}\omega\rVert^2$\;
				\For {$c = 1$ \textbf{to} $J$}
				{
					\For {$i = 2$ \textbf{to} $M$}
					{
						$\xi_1 = -\frac{1}{\sigma^2}\lVert\mathbf{y}_k-\mathbf{h}_dx_i + \underline{\mathbf{f}}\boldsymbol{\Lambda}_cx_i \rVert^2$\;
						$\xi_2 =  \max \left\{\Delta_i, \xi_1\right\}+  \ln(1+\exp(-\lvert  \xi_1 - \Delta_i \rvert))$\;
						$\Delta_i = \xi_2$\;
					}
				}
				$\vartheta_k = \ln(L^2) - \ln((K-L)^2) + \Delta_i - \Delta_p$\;	
			}
			\KwRet $\boldsymbol{\vartheta}$				
		\end{algorithm}
		
		\begin{algorithm}[b!]  
			\label{TIM_LLR}
			\SetAlgoLined
			\caption{Mapping Rule-aided Info. Slot Selection} 
			\DontPrintSemicolon
			\SetKwInOut{Parameter}{Require}
			\SetKwInOut{Input}{Ensure}
			\Parameter{$\boldsymbol{\vartheta}$, $\mathcal{T}$, $\eta_r$, $K$, $L$}
			\Input{$\mathbf{t}_\alpha$ belongs to the legitimate set $\mathcal{T}$ and ${\mathbf{t}_\alpha^\iota}$ is the $\iota$-th element of $\mathbf{t}_\alpha$.}
			
			$\vartheta = \boldsymbol{\vartheta}_k$\;
			$	\mathbf{b} = \mathbf{0}_{2^{\eta_r} \times 1}$\;
			\For {${\alpha} = 1$ \textbf{to} $2^{\eta_r}$}
			{
				\If{$L > 1$}{
					\For {${\iota} = 1$ \textbf{to} $L$}		
					{	
						$a$ = ${\mathbf{t}_\alpha^\iota}$\;
						$\mathbf{b}_{\alpha} \mathrel{+}= \boldsymbol{\vartheta}_a$\;
					}	
					
				}
				\ElseIf{L = 1} {
					$a$ = ${\mathbf{t}_\alpha}$\;
					$\mathbf{b}_{\alpha} =  \boldsymbol{\vartheta}_a$\;
				}
			}
			$\mathbf{t}_{\alpha} = \arg \max_{\alpha}\left\{\mathbf{b}_{\alpha}\right\}$\;
			\KwRet $\boldsymbol{\hat{\tau}_{\mathbf{t}_\alpha}}$			
		\end{algorithm}			
		Since the ML detector searches for all possible mapping rules and the corresponding symbols, the decoding complexity increases as $K$, $L$, and $M$ grow. The computational complexity of the ML detector is approximately $\mathcal{O}(2^{\eta_r}JM^L)$ in terms of complex multiplications. Therefore, we propose an LLR-based detector for our scheme to reduce the decoding complexity while maintaining the reliability of decoding at PRx.
		
		Let us define two vectors for the information and power stages from $\boldsymbol{\Psi}_k$ to ease the exposition. To describe the power stage, we can represent $\boldsymbol{\Psi}_k$ as $\boldsymbol{\Omega} = \begin{bmatrix}e^{-j \phi_p}& 0 &e^{-j \phi_p} \end{bmatrix}$, and for the information stage, $\Lambda_c = \begin{bmatrix}\psi_1& 0 &\theta_c \end{bmatrix}$. Now, we can write the expression representing the logarithm ratio of the a posteriori probabilities of the symbols, taking into account that their values may indicate the selection of either information or power signal. The expression is as follows:
		\begin{equation}
		 	\vartheta_k =
		 	\ln \left(\frac{\sum_{j=1}^{{J}}\sum_{i=1}^{M} \Pr\left(s_k=x_i, \psi_3 = \theta_c \mid \mathbf{y}_k\right)}{\Pr\left(s_k=\omega, \psi_3 = \phi_p \mid \mathbf{y}_k\right)}\right).
		 	\label{eq:LLR1}
		 \end{equation}
		 
		 A higher value of $\vartheta_k$ in \eqref{eq:LLR1} indicates a higher likelihood of information being transmitted at time instant $k$.  Applying Bayes' formula, with $\Sigma_{i=1}^{M} \Pr\left(s_k= x_i\right) = \Sigma_{c=1}^{J} \Pr\left(s_k=\theta_c\right) =L \slash K$ and $\Pr(s_k=\omega ) = \Pr\left(\psi_3 = \phi_p \right) = (K-L)\slash K$, \eqref{eq:LLR1} can be expanded as: 
		\begin{align}
				\vartheta_k =  
				&\ln\left(\sum_{j=1}^{{J}}\sum_{i=1}^{M} \exp \left(-\frac{1}{\sigma^2}\lVert\mathbf{y}_k-\mathbf{h}_d x_i + \underline{\mathbf{f}}\boldsymbol{\Lambda}_cx_i\rVert^2\right)\right) \nonumber\\
				&-\ln \left( \exp \left(-\lVert\mathbf{y}_k-\mathbf{h}_d \omega + \underline{\mathbf{f}}\boldsymbol{\Omega}\omega \rVert^2\right)\right)\\
				& + \ln \left({L^2}\right) -  \ln \left({(K - L)^2}\right)\nonumber.			
		\end{align}
			\begin{table}[!t]
			\centering
			\caption{MAPPING RULE of $\boldsymbol{{\tau}_{\mathbf{t}_\alpha}}$ for $K = 4$ and $ L = 2$.}
			\label{tab:TIM_Mapping}
			\resizebox{0.6\columnwidth}{!}{%
				\begin{tabular}{|c|c|c|}
					\hline
					\vspace{-0.1cm} &\vspace{-0.1cm} & \vspace{-0.1cm}\\
					\textbf{Bits} & $\boldsymbol{{\tau}_{\mathbf{t}_\alpha}}$ &$\mathcal{T}$\\ \hline\hline
					\vspace{-0.1cm}  & \vspace{-0.1cm}&\\
					00           & $\begin{bmatrix}1 & 0 & 1 & 0\end{bmatrix}$ & $\mathbf{t}_{1} = \begin{bmatrix}	1 & 3\end{bmatrix}$\\
					\vspace{-0.1cm}  & \vspace{-0.1cm}&\\
					01            & $\begin{bmatrix}1 & 0 & 0 & 1\end{bmatrix}$ &$\mathbf{t}_{2} = \begin{bmatrix}	1 & 4\end{bmatrix}$\\
					\vspace{-0.1cm} & \vspace{-0.1cm}&\\
					10           & $\begin{bmatrix}0 & 1 & 0 & 1\end{bmatrix}$ &$\mathbf{t}_{3} = \begin{bmatrix}	2 & 4\end{bmatrix}$\\
					\vspace{-0.1cm}  & \vspace{-0.1cm}&\\
					00            & $\begin{bmatrix}0 & 1 & 1 & 0\end{bmatrix}$ &$\mathbf{t}_{4} = \begin{bmatrix}	2 & 3\end{bmatrix}$\\
					\vspace{-0.1cm} & \vspace{-0.1cm}&\\
					X            & $\begin{bmatrix}1 & 1 & 0 & 0\end{bmatrix}$ &-\\
					\vspace{-0.1cm} &\vspace{-0.1cm} & \vspace{-0.1cm}\\
					X            & $\begin{bmatrix}0 & 0 & 1 & 1\end{bmatrix}$ &-\\ \hline
					
				\end{tabular}
			}
		\end{table}
		
		\begin{figure}[b!]
			\centering
			\subfloat[\label{fig:RIS_Harvested}]{\includegraphics[width=0.8\linewidth]{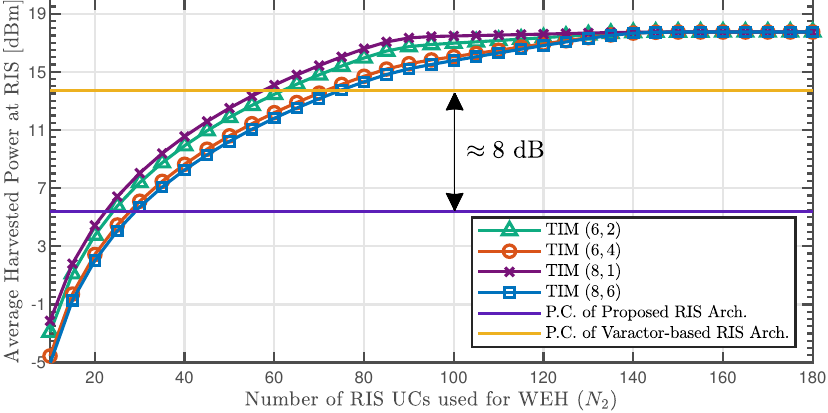}}
			\vspace{\intextsep}\vspace{0.1cm}
			\subfloat[\label{fig:EH_Harvested}]{\includegraphics[width=0.8\linewidth]{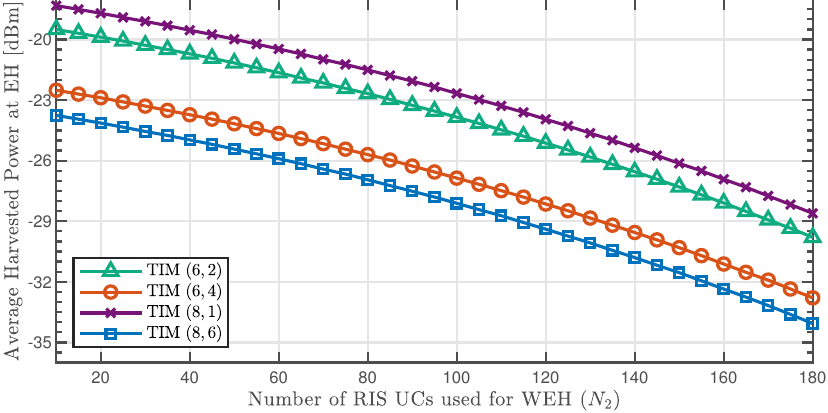}}
			\label{fig:Harvested}
			\caption{Average harvested DC power \eqref{eq:Avg_Harvested} with respect to $N_2$ UCs  for (a) standalone RIS and (b) EH.}
		\end{figure}

		Then, recursion initializes with $\Delta_i$ = $-\frac{1}{\sigma^2} \lVert\mathbf{y}_k-\mathbf{h}_d x_1 + \underline{\mathbf{f}}\boldsymbol{\Lambda}_1x_1\rVert^2 $. To avoid numerical overflow, we utilize the following identity: $
		\ln \left(e^{\xi_1}+e^{\xi_2}+\cdots+e^{\xi_\chi}\right) =\ln \left(e^{\Delta_i}+e^{\xi_\chi}\right) =\max \left\{\Delta_i, \xi_\chi\right\}+\ln \left(1+e^{-\left|\xi_\chi-\Delta_i\right|}\right)$ \cite{Basar2013} at each iteration in which $\xi_{\chi} = \lVert\mathbf{y}_k-\mathbf{h}_d x_i + \underline{\mathbf{f}}\boldsymbol{\Lambda}_jx_i\rVert^2$ by setting $c = 1$ to $J$ and $i = 2$ to $M$. Since there is one possible power signal  once 	$\Delta_p$ = $-\lVert\mathbf{y}_k-\mathbf{h}_d \omega + \underline{\mathbf{f}}\boldsymbol{\Omega}\omega\rVert^2$ is evaluated and keep the same for each iteration. Afterward, LLR value is obtained by  Algorithm 1 for each time instant $k$ and they are collected in a vector which is given by
		\begin{equation}
			\boldsymbol{\vartheta} = \begin{bmatrix}\vartheta_1 & \vartheta_2 &\hdots &\vartheta_k\end{bmatrix}.
		\end{equation}
			\begin{figure*}[t!]
			\centering
			\subfloat[\label{fig:BER_PTx}]{\includegraphics[width=0.33\linewidth]{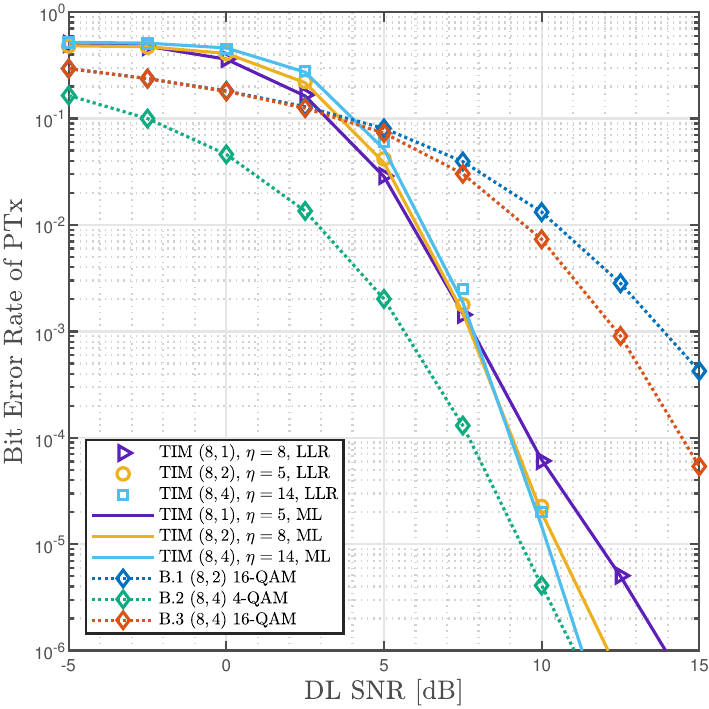}}\hfill
			\subfloat[\label{fig:BER_RIS}]{\includegraphics[width=0.33\linewidth]{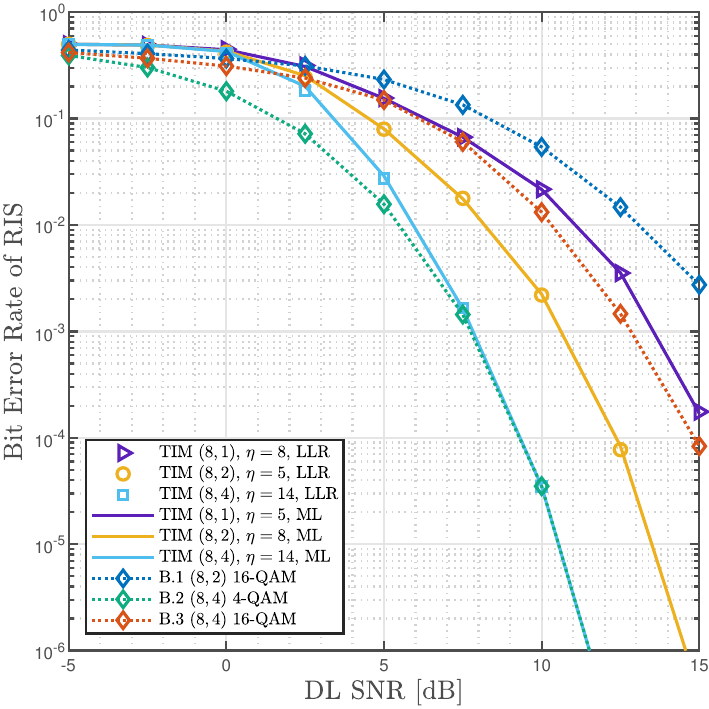}}\hfill
			\subfloat[\label{fig:PH_BER}]{\includegraphics[width=0.33\linewidth]{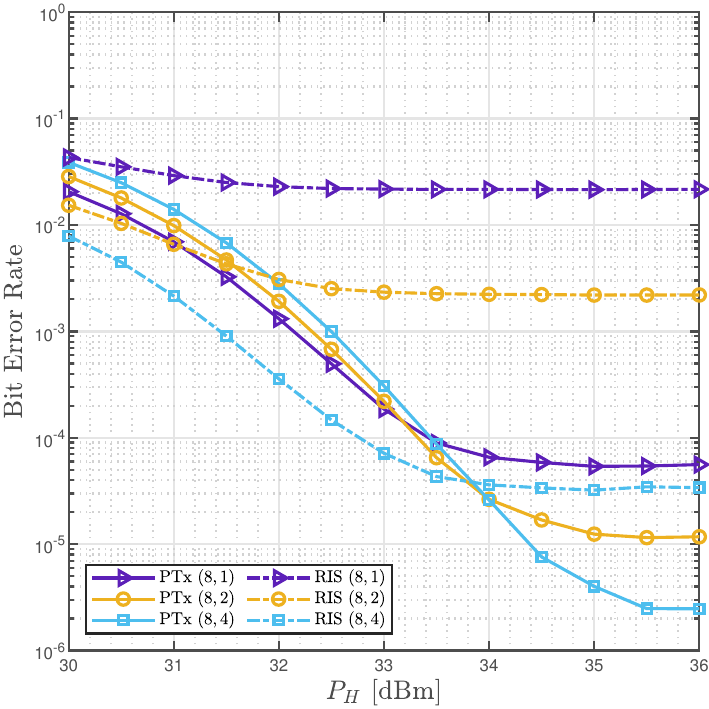}}
			\vspace{0.1cm}
			\caption{BER of the proposed scheme for various TIM setups: (a) PTx, (b) RIS, and (c) PTx and RIS for varying $P_H$ values.}
			\label{fig:BER}
		\end{figure*}	
		Finally, the receiver decides on the time indices of $L$ information slots $\mathbf{t}_\alpha$ based on maximum LLR values with the aid of $\boldsymbol{\vartheta}$ and estimates $\boldsymbol{\hat{\tau}_{\mathbf{t}_\alpha}}$. It can be seen that the proposed LLR detector simplifies decoding complexity to around $\mathcal{O}(K(JM + 1))$ in terms of complex multiplications.

		The mapping rule for $(K = 4, L = 2)$ is demonstrated in Table \ref{tab:TIM_Mapping}. When selecting $(4, 2)$, there are $6$ possible mappings. Nevertheless, only $2^{\left\lfloor \log_2 (6) \right\rfloor} = 4$ legitimate mappings can be chosen, as there are four possible bit combinations. Therefore, it is likely to decide on an illegitimate set with Algorithm 1. 
			
		Assuming that $\hat{L}$ is the number of estimated information slots in $\boldsymbol{\hat{\tau}_{\mathbf{t}_\alpha}}$, in cases where $\hat{L} > L$ or $\hat{L} < L$, or even when $\hat{L} = L$, the receiver can detect $\boldsymbol{{\tau}_{\mathbf{t}\alpha}}$, which is not specified in the mapping rule, such that the detection process will fail. Thus, we present an LLR detector that operates in cooperation with the mapping rule to prevent the detection of illegitimate sets. Hence, regarding the potential scenario where the PRx might choose $\boldsymbol{{\tau}_{\mathbf{t}\alpha}}$ that does not belong to the mapping rule, Algorithm 2 is provided to compute the maximum LLR sum for determining time-indices $\mathbf{t}_\alpha$ from legitimate set $\mathcal{T}$ \cite{Basar2013}. Ultimately, to estimate $(\mathbf{\hat{x}}_n, \hat{\psi}_3)$ for selected time instants (information slots), the receiver applies an ML detector, which can be expressed as:
		\begin{equation}
			\left(\mathbf{\hat{x}}_n, \hat{\psi}_3\right) = \arg \min_{\substack{{\mathbf{x}_n \in \mathcal{S}}\\ {\psi_3 \in \mathcal{J^{\prime}}}}}  \sum_{\alpha=1}^{L}\lVert\mathbf{y}_\alpha-\mathbf{h}_d s_\alpha + \underline{\mathbf{f}}\boldsymbol{\Psi}_\alpha s_\alpha\rVert^2.
		\end{equation}
		where $\mathbf{y}_\alpha , s_\alpha , \boldsymbol{\Psi}_\alpha$ corresponds to the estimated time indices in  $\boldsymbol{\hat{\tau}_{\mathbf{t}_\alpha}}$ computed by Algorithm 2.
		
		\section{Numerical Results}
		In this section, we evaluate the performance of our proposed scheme in terms of average harvested DC power and bit error rate (BER). For all channels, a Rician factor $\kappa_p = 5$ is considered, and the 3GPP Indoor hotspot (InH) path loss model is employed, which is expressed as follows \cite{3GPP_PL}:
		\begin{equation}
			\nu_p(d) = 32.8 + 16.9 \log _{10} (d) + 20 \log _{10} (f),
		\end{equation}
		where $f$ is the carrier frequency, and $d$ refers to the distance, with $d \in \{5, 10, 14\}$ for the PTx-RIS, RIS-PRx (EH), and PTx-RIS-PRx (EH) links, respectively. Unless otherwise specified, $4$-QAM is employed at PTx, and $M_R = 4$, and $N_1 = 60$ are used in the simulations. The simulation parameters are presented in Table \ref{tab:Sim}.
		
			\begin{table}[b!]
			\centering
			\caption{SIMULATION PARAMETERS}
			\label{tab:Sim}
			\resizebox{0.8\columnwidth}{!}{%
				\begin{tabular}{ccccccccc}
					\cline{2-9}
					\multicolumn{1}{c|}{}                      & \multicolumn{4}{c|}{\textbf{Parameter}}                              & \multicolumn{4}{c|}{\textbf{Value}}                    \\  \cline{2-9} 
					\vspace{-0.09cm}     &                \vspace{-0.09cm}                  &        \vspace{-0.09cm}           \\ \hline
					\multicolumn{1}{|c|}{\multirow{2}{*}{\textbf{PTx}}} & \multicolumn{4}{c|}{$P_L, P_H $}                            & \multicolumn{4}{c|}{$30$ dBm, $34$ dBm}       \\
					\multicolumn{1}{|c|}{}                     & \multicolumn{4}{c|}{$f$}                                    & \multicolumn{4}{c|}{$2$ GHz}                  \\ \hline
					\vspace{-0.09cm}     &                \vspace{-0.09cm}                  &        \vspace{-0.09cm}           \\ \hline
					\multicolumn{1}{|c|}{\multirow{4}{*}{\textbf{RIS}}} & \multicolumn{4}{c|}{$N$, $N_\text{cb}$}                     & \multicolumn{4}{c|}{$256, 4$}                 \\
					\multicolumn{1}{|c|}{}                     & \multicolumn{4}{c|}{$P_\text{cb}$, $P_\text{dyn}$}          & \multicolumn{4}{c|}{$50$ $\mu$W , $40$ $\mu$W \cite{Petrou2022}} \\
					\multicolumn{1}{|c|}{}                     & \multicolumn{4}{c|}{$P_\text{switch}$, $P_\text{varactor}$} & \multicolumn{4}{c|}{$1 \mu$W \cite{rossanese2022}, $ 0$ W \cite{wang2024reconfigurable}}         \\
					\multicolumn{1}{|c|}{}   & \multicolumn{4}{c|}{$\rho$, $\mathcal{P}_{\text{on}}$, $\mathcal{P}_\text{sat}$} & \multicolumn{4}{c|}{$0.75$, $150$ $\mu$W, $70$ mW} \\ \hline
					\vspace{-0.09cm}     &                \vspace{-0.09cm}                  &        \vspace{-0.09cm}           \\ \hline
					\multicolumn{1}{|c|}{\textbf{EH}} & \multicolumn{4}{c|}{$\rho$, $\mathcal{P}_{\text{on}}$, $\mathcal{P}_\text{sat}$} & \multicolumn{4}{c|}{$0.75$, $50$ $\mu$W, $0.1$ mW} \\ \hline
					
				\end{tabular}%
			}
		\end{table}

		We aim to demonstrate the proposed standalone RIS architecture in terms of average harvested DC power and the number of $N_2$ UCs required to meet the condition \eqref{eq:zeRIS}. Fig. \ref{fig:RIS_Harvested} indicates that the minimum number of $N_2$ UCs required for a standalone RIS architecture is approximately two times fewer for RF switches compared to varactors. This difference arises as varactors are driven by power-hungry DACs that make accomplishing a standalone operation challenging, resulting in roughly 8 dB more power consumption than RF switches. As expected, reserving more time slots for the power stage in the TIM scheme increases the average harvested DC power, as illustrated in Figs. \ref{fig:RIS_Harvested} and \ref{fig:EH_Harvested}. Importantly, around $N_2 = 140$, the CLC rectenna reaches saturation region for all TIM schemes. Herein, it is crucial to mention the parameters of the CLC rectenna model can be reconfigured according to $N_2$ UCs to leverage the benefits effectively. Furthermore, allocating more $N_2$ UCs at the RIS reduces the average harvested DC power at the EH, as the total number of reflectors, which is $N - N_2$ UCs, decreases. Since $N_2 = 35$ satisfies the standalone condition \eqref{eq:zeRIS} for all schemes in Fig. \ref{fig:RIS_Harvested}, we exploit this in our simulations to ensure that RIS performs its functions by WEH in the SR environment.

		In Figs. \ref{fig:BER_PTx} and \ref{fig:BER_RIS}, we demonstrate the BER performance of our scheme for ML and LLR detectors with respect to direct link (DL) signal-to-noise ratio (SNR), $\gamma_{\text{DL}} = \nu_{h_d}(d) / \sigma^2$. It can be observed from Figs. \ref{fig:BER_PTx} and \ref{fig:BER_RIS} that BER performances of the ML and LLR detectors align with one another. However, the LLR detector reduces the complexity by approximately $86\%$ compared to the ML detector in terms of complex multiplications, considering the TIM $(8, 2)$ scheme.  Additionally, based on the setup in \cite{Yang2024}, we compare our scheme with benchmarks. We assume that the benchmark schemes maintain RIS information unchanged for $L$ information slots for a block size of $K=8$, without utilizing TIM at the PTx. Thus, the benchmark schemes convey $\eta_m = L\log_2(M)$ bits and $(K-L)$ power signals in a given block. In Fig. \ref{fig:BER_PTx}, the benchmark scheme B.2  outperforms our proposed scheme. Nevertheless, when considering the $(8, 2)$ case, since our scheme exploits six time slots for power transfer while the benchmark B.2 uses four time slots, the average harvested DC power in a block size of $K$ should be higher than that of B.2. If the benchmark utilizes six time slots (B.1) for power transfer, all TIM schemes show their superiority against it. Thus, modulating bits in the time domain rather than the signal domain enhances the BER performance of the SR system. In Fig. \ref{fig:BER_RIS}, the proposed scheme outperforms benchmark B.1 in terms of BER of RIS information. Despite B.3 having a spreading gain four times that of the TIM $(8, 1)$ case, their performances are close due to the fact that PTx and RIS information are coupled. Therefore, increasing the modulation order $M$ of PTx has a detrimental impact on RIS information. Additionally, as the DL SNR increases, the BER of B.2 and TIM $(8, 4)$ cases become indistinguishable. Moreover, the BER performance of the RIS improves with increasing spreading gain $\delta$, which also enhances the BER of PTx in the high SNR region, as observed in Fig. \ref{fig:BER_PTx}. 
		
		To emphasize the effect of $P_H$ on decoding performance in the LRR detector, the BER for PTx and RIS information is plotted in Fig. \ref{fig:PH_BER} with $\gamma_{\text{DL}} = 10$ dB. As $P_H$ increases, the distinction between information and power stages becomes more accurate, leading to a better estimation of time-index bits in the LLR detector. Interestingly, the impact of $P_H$ on the BER of RIS information is highly coupled with the spreading factor $\delta$. Therefore, as $\delta$ improves, increasing $P_H$ becomes more meaningful for achieving better BER performance. Furthermore, the BER of PTx and RIS information saturates after a specific SNR value in all TIM schemes, indicating that further increasing $P_H$ will not affect decoding performance.

		\section{Concluding Remarks}
		The coexistence of cellular and IoT devices in next-generation communication frameworks becomes more power and bandwidth-hungry as the number of connected devices increases. Therefore, delivering power and information in a mutualistic manner is of paramount importance for designing next-generation networks. This paper presents a novel low-power, standalone RIS architecture that serves as a BD in an SR environment. In this setup, the PTx utilizes a TIM scheme to transfer information to the PRx and power to the RIS and EH, respectively. We propose an LLR-based detector to reduce computational complexity at the PRx and investigate the proposed scheme in terms of average harvested DC power and BER performance while ensuring the necessary conditions for being a standalone RIS. Our results demonstrate that a standalone RIS can establish a mutualistic symbiosis with the PTx, and the proposed system can achieve reliable communication while incorporating power transfer into its transmission strategy without demanding additional power resources.

		\bibliographystyle{IEEEtran}
		\bibliography{TIMRISref}
		
	\end{document}